\documentclass[prb,aps]{revtex4}
\newcommand{\dlangle}{\langle\langle}
\newcommand{\drangle}{\rangle\rangle}
\newcommand{\up}{\uparrow}
\newcommand{\dn}{\downarrow}
\newcommand{\down}{\downarrow}
\begin{document}

\title{Spin waves in ultrathin ferromagnetic overlayers}

\author{L. H. M. Barbosa} 
\affiliation{Departamento de F\'\i sica, Universidade Federal Fluminense,
24210-340 Niter\'oi, Brazil} 
\author{A. T. Costa Jr.}
\affiliation{Departamento de Ci\^encias Exatas, Universidade Federal de Lavras,
37200-000 Lavras, MG, Brazil}
\author{J. Mathon} 
\affiliation{Department of
Mathematics, City University, London EC1V 0HB, U.K.}
\author{R. B. Muniz}
\affiliation{Departamento de F\'\i sica, Universidade Federal Fluminense,
24210-340 Niter\'oi, Brazil}

\date{\today}

\begin{abstract} 
  
  The influence of a non-magnetic metallic substrate on the spin wave
  excitations in ultrathin ferromagnetic overlayers is investigated for
  different crystalline orientations. We show that spin wave dumping in these
  systems occur due to the tunneling of holes from the substrate into the
  overlayer, and that the spin wave energies may be considerably affected by
  the exchange coupling mediated by the substrate.

\end{abstract}

\pacs{PACS numbers: 75.50.Fr, 75.30.Et, 75.50.Rr}

\maketitle

\section{Introduction}

Recent advances in materials growth techniques and precise control of
deposition processes have enabled production of multilayers with excellent
interfacial quality.\cite{paggel,unguris} Presently, it is possible to grow
ultrathin magnetic films on a substrate, regulating the film thickness very
accurately. Epitaxial films with very well-defined thicknesses, showing
virtually no layer thickness fluctuation over macroscopic distances, have been
fabricated.\cite{paggel} The ability to control film thicknesses with such
accuracy, together with the freedom of choosing different substrates, with
distinct crystalline orientations, broaden the spectrum of magnetic responses,
and make these systems highly attractive for technological applications.

The magnetic behavior of ultrathin films is strongly affected by spin wave
excitations. Hence, the study of spin waves in these structures is important
for understanding their magnetic properties and characteristics. In fact, spin
waves in ultrathin films have been extensively studied, both experimentally
and theoretically.\cite{Mills,3,Swr} At low temperatures, for instance, the
magnetization reduction is basically controlled by long-wavelength spin waves.
Thus, measurements of $M(T)$ can provide useful information about spin wave
excitations in low-dimensional magnetic strucutures. Generally, the energy of
a long-wavelength spin wave propagating with wave vector $\vec{q}$ is given by
$E=D(\hat{q}) q^2$, where $D(\hat{q})$ is the exchange stiffness constant. In
a relatively thin magnetic layer, spin waves are excited with wave vectors
$\vec{q_\parallel}$ parallel to the layer, and for certain crystalline
orientations, the spin wave energies may depend much upon the direction of
$\vec{q_\parallel}$ \cite{Tovar}, due to lattice anisotropies.

It has also been shown \cite{phan} that spin wave lifetimes in ultrathin
ferromagnetic films can be substantially affected by the presence of a
non-magnetic substrate. By considering a monolayer of a strong ferromagnet on
a surface of a non-interacting metallic substrate, Mathon et al. \cite{phan}
showed that spin waves in the overlayer become critically damped. Such
behavior has been attributed to the decaying of spin waves into electron-hole
pairs due to the tunneling of holes from the substrate into the magnetic
overlayer. The substrate also affects the spin wave energy, and modify the
region in $q_\parallel$-space where it follows a quadratic behavior.

Here we pursue these ideas and investigate the influence of a non-magnetic
substrate in spin wave excitations in ultrathin ferromagnetic overlayers for
different crystalline orientations. We consider a monolayer of a strong
ferromagnet both on (100) and (110) surfaces of a semi-infinite nonmagnetic
metallic substrate. By artificially reducing the electron hopping between the
substrate and the overlayer we explicitly demonstrate that the spin wave
dumping is really due to the tunneling of holes from the substrate into the
overlayer, as previously pointed out by Mathon et al.\cite{phan}. For the
(110) overlayer we have found that the spin wave energy depend upon the
$\vec{q_\parallel}$ direction, but such dependence is much less pronounced than
previously found for unsuported monolayers. This is due to the enhancement of
the exchange coupling between the local moments in the overlayer mediated by
the substrate, which strongly affects the spin wave energies.

This paper is organized as follows: in section 2 we briefly review the theory
we have used to calculate the spin wave energies and lifetimes in overlayers.
In section 3 we present our results and discussions, and finally, in section
4, we draw our main conclusions.
 
\section{Transverse spin susceptibility}
The spin-wave spectrum of itinerant ferromagnets can be obtained from the
dynamical transverse spin susceptibility
\begin{equation}
\chi^{+-}(\vec{q},\omega) = \sum_{j} \int_{-\infty}^{\infty}dt
e^{-i\omega t}e^{-i\vec{q}.\left(\vec{R_i}-\vec{R_j}\right)}
\chi^{+-}_{ij}(t)\, 
\label{eqt1}
\end{equation}
where $\chi^{+-}_{ij}(t)=\dlangle S^{+}_i(t);S^{-}_j(0) \drangle$ is the
time-dependent transverse spin susceptibility in real space, given by the
two-particle Green's function
\begin{equation}
\chi^{+-}_{ij}(t)=-i\Theta(t)\langle\left[ S^{+}_{i}(t),S^{-}_{j}(0)\right]
\rangle \, .
\label{eqt2}
\end{equation}
Here, $S_i^{+}$ ($S_i^{-}$) is the spin raising (lowering) operator at site
$i$, the brackets denote a commutator, $\langle ...\rangle$ represents the 
thermodinamical average which at zero temperature reduces to the ground-state 
expectation value, and $\Theta(t)$ is the usual step function given by
\[ 
\Theta(t) = \left\{ \begin{array}{ll}
0 & \mbox{if $t<0$} \\
1 & \mbox{if $t>1$}
\end{array}
\right. \, . 
\] 

The exact calculation of $\chi^{+-}_{ij}(t)$ involves the solution of an
infinite set of coupled equations which in general is not possible to find.
However, the random phase approximation (RPA) provides an useful decoupling
scheme which allows one to solve such equations and to obtain the spin wave
energies rather accurately. To find an expression for $\chi^{+-}_{ij}(t)$
within the RPA, we follow references [\onlinecite{Wolfran,Mills2}], and define
the spin operators
\begin{eqnarray}
S^{+}_{ij}=a^\dagger_{i\uparrow}a_{j\downarrow}
\nonumber\\
S^{-}_{ij}=a^\dagger_{i\downarrow}a_{j\uparrow}\, ,
\label{genop}
\end{eqnarray}
where $a^\dagger_{i\sigma}$ ($a_{i\sigma}$) creates (destroys) an electron 
with spin $\sigma$ at site $i$. Using $S^{\pm}_{ij}$, one may define a 
generalized susceptibility 
\begin{equation}
\chi^{+-}_{ijkl}(t) = \dlangle S^{+}_{ij}(t);S^{-}_{kl}(0)\drangle\, .
\label{gensus}
\end{equation}
Clearly, the transverse spin susceptibility we wish to calculate is
$\chi^{+-}_{ij}(t) = \chi^{+-}_{iijj}(t)$. In order to calculate it, we 
consider that the electronic structure of the system is described by a simple 
one-band Hubbard Hamiltonian
\begin{equation}
H = \sum_{ij\sigma} t_{ij}a^\dagger_{i\sigma}a_{j\sigma} + 
\sum_{i\sigma} \epsilon_{i} n_{i\sigma} + 
\sum_{i} U_{i} n_{i\uparrow} n_{i\downarrow},
\label{Hub-H} 
\end{equation}
where $t_{ij}$ is the hopping integral between sites $i$ and $j$ ($t_{ii}=0$), 
$\epsilon_{i}$ is an atomic energy level, $U_{i}$ represents the effective 
Coulomb interaction between two electrons on the same site $i$, and 
$n_{i\sigma}=a^\dagger_{i\sigma}a_{i\sigma}$ is the corresponding electronic 
occupation number.

In this case, $\chi^{+-}_{ijkl}(t)$ obeys the following equation of motion:
\begin{eqnarray}
i\hbar\frac{d\chi^{+-}_{ijkl}(t)}{dt}=\delta(t)\langle a_{i\up}^\dagger 
a_{l\up}\delta_{jk}-a_{k\dn}^\dagger a_{j\dn}\delta_{il}\rangle + 
\sum_n\left[ t_{jn}\chi^{+-}_{inkl}(t)-t_{ni}\chi^{+-}_{njkl}(t)\right]+
\nonumber\\
+ U_j\dlangle \left[a^\dagger_{i\up}a_{j\dn}a^\dagger_{j\up}a_{j\up}
\right](t);S^-_{kl}(0)\drangle+U_i\dlangle \left[a^\dagger_{i\dn}a_{i\dn}
a^\dagger_{i\up}a_{j\dn}\right](t);S^-_{kl}(0)\drangle
\label{eqmotion}
\end{eqnarray}

The appearance of higher order Green functions leads to an infinite chain of
coupled equations for $\chi^{+-}_{ijkl}(t)$ that can be decoupled using the
RPA which consists of replacing
\begin{eqnarray}
a^\dagger_{i\sigma}a_{j\sigma'}a^\dagger_{k\xi}a_{l\xi'}\approx
\langle a^\dagger_{i\sigma}a_{j\sigma'}\rangle a^\dagger_{k\xi}a_{l\xi'} -
\langle a^\dagger_{i\sigma}a_{l\xi'} \rangle a^\dagger_{k\xi}a_{j\sigma'} +
\langle a^\dagger_{k\xi}a_{l\xi'} \rangle a^\dagger_{i\sigma}a_{j\sigma'} -
\langle a^\dagger_{k\xi}a_{j\sigma'}\rangle a^\dagger_{i\sigma}a_{l\xi'}\, .
\label{RPA}
\end{eqnarray} 
Here, the expectation values are evaluated in the Hartree-Fock ground state,
where the dynamics of the two spin projections are treated independently. As a
result, the average of products of two operators associated with opposite spins
vanishes.  Therefore, one obtains

\begin{eqnarray}
i\hbar\frac{d\chi^{+-}_{ijkl}(t)}{dt}= \delta(t)\langle a_{i\up}^\dagger 
a_{l\up}\delta_{jk}-a_{k\dn}^\dagger a_{j\dn}\delta_{il}\rangle+
\sum_{m,n}\left(\delta_{im}t_{jn}-\delta_{jn}t_{mi}\right)\chi^{+-}_{mnkl}(t)+
\nonumber\\
+\sum_{m,n}\delta_{mn}U_m\left(\delta_{im}\langle a^\dagger_{i\dn}a_{j\dn}
\rangle-\delta_{jm}\langle a^\dagger_{i\up}a_{j\dn}\rangle\right)\chi^{+-}
_{mnkl}(t)\, .
\label{eqmRPA}
\end{eqnarray}

The Fourier transform of equation \ref{eqmRPA} then reads
 
\begin{equation}
\hbar\omega\chi^{+-}_{ijkl}(\omega)=D_{ijkl} + 
\sum_{mn}\left[K_{ijmn}\chi^{+-}_{mnkl}(\omega) +
J'_{ijmn}\chi^{+-}_{mnkl}(\omega)+J_{ijmn}\chi^{+-}_{mnkl}(\omega)\right]\, ,
\label{eqmRPAm}
\end{equation}
where $\hat{D}$, $\hat{K}$, $\hat{J'}$ and $\hat{J}$ are four-indices 
matrices defined by
\begin{eqnarray}
D_{ijkl} = \langle a^\dagger_{i\up}a_{l\up}\delta_{jk} - 
a^\dagger_{k\dn}a_{k\dn}\delta_{il}\rangle \nonumber\\
K_{ijkl} = \delta_{ik}t_{jl} - \delta_{jl}t_{ki}\nonumber\\
J_{ijkl} = \delta_{kl}U_k\left(\delta_{ik}\langle a^\dagger_{i\dn}a_{j\dn}
\rangle-\delta_{jk}\langle a^\dagger_{i\up}a_{j\up}\rangle\right)\nonumber\\
J'_{ijkl} = \delta_{ik}\delta_{jl}\left(U_j\langle n_{j\up}\rangle-
U_i\langle n_{i\dn}\rangle\right)\, .
\label{matrices}
\end{eqnarray}
The matrix elements of the product of two of such matrices is given by 
$\left(\hat{A}\hat{B}\right)_{ijkl} = \sum_{mn}A_{ijmn}B_{mnkl}$. Thus, 
we may rewrite equation \ref{eqmRPAm} in matrix form as 
\begin{equation}
\hbar\omega\hat{\chi}^{+-}(\omega) = \hat{D}+(\hat{K}+\hat{J}+\hat{J'})
\hat{\chi}^{+-}(\omega)\, .
\label{eqmRPAfin}
\end{equation}
This equation may be also rewritten as 
\begin{equation}
\hat{\chi}^{+-}(\omega) =\hat{\chi}^{HF}(\omega) + 
\hat{\chi}^{HF}(\omega)\hat{P}\hat{\chi}^{+-}(\omega)\, ,
\label{RPAHF}
\end{equation}
where $\hat{P}=\hat{D}^{-1}\hat{J}$, and $\hat{\chi}^{HF}(\omega)$ represents
the susceptibility $\hat{\chi}^{+-}(\omega)$ calculated within one-electron 
theory, i.e., in the Hartree-Fock approximation. $\hat{\chi}^{HF}(\omega)$ 
satisfies the following equation:

\begin{equation}
\hbar\omega\hat{\chi}^{HF}(\omega) = \hat{D}+(\hat{K}+\hat{J'})
\hat{\chi}^{HF}(\omega)\, .
\label{eqmHF}
\end{equation}
Therefore, by using equations \ref{matrices} we find 
\begin{equation}
\chi^{+-}_{ijkl}(\omega) = \chi^{HF}_{ijkl}(\omega) - \sum_m
\chi^{HF}_{ijmm}(\omega)U_m\chi^{+-}_{mmkl}(\omega)\, .
\label{RPAHFfin}
\end{equation}

The dynamic susceptibility $\chi^{+-}_{ij}(t) = \chi^{+-}_{iijj}(t)$ is then
given by:
\begin{equation}
\chi^{+-}_{ij}(\omega) = \chi^{HF}_{ij}(\omega) - \sum_m
\chi^{HF}_{im}(\omega)U_m\chi^{+-}_{mj}(\omega)\, .
\label{fin}
\end{equation}

In the HF approximation, $\uparrow$- and $\downarrow$-spin electrons are
independent, and an electron with spin $\sigma$ at site $i$ is subjected 
to the HF potential
\begin{equation}
\epsilon_{i\sigma} = \epsilon_{i} + U_{i} \langle n_{i-\sigma}\rangle\, . 
\end{equation}
It follows that $\chi^{HF}_{ij}(\omega)$ can be expressed in terms of the HF
one-electron propagators as
\begin{equation}
\chi^{HF}_{ij}(\omega) = -\langle a^\dagger_{i\up}(t)a_{j\up}\rangle G_{ij}^{
\down}(t) + \langle a^\dagger_{j\down}a_{i\down}(t)\rangle G_{ij}^{\up *}(t)
\, ,
\label{chiHF}
\end{equation}
where $G_{ij}^{\sigma}(t)$ is the time-dependent one-particle retarded Green
function for electrons with spin $\sigma$, connecting sites $i$ and $j$, 
defined by
\begin{equation}
G_{ij}^{\sigma}(t)=-\frac{i}{\hbar}\theta(t)
\langle\left\{ a_{i\sigma}(t),a_{j\sigma}^\dagger\right\}\rangle\, ,
\label{opgf}
\end{equation}
where the braces denotes an anticommutator.
The correlation functions in equation \ref{chiHF} can be written in terms of 
the one-particle Green functions as

\begin{eqnarray}
\langle a_{i\sigma}^\dagger(t)a_{j\sigma}\rangle = \frac{1}{\pi}\int d\omega 
f(\omega)\, \Im m\, G_{ji}^{\sigma}(\omega)\, e^{i\omega t}\, ,
\nonumber\\
\langle a^\dagger_{j\sigma}a_{i\sigma}(t)\rangle = -\frac{1}{\pi}\int d\omega 
f(\omega)\, \Im m\, G_{ij}^{\sigma}(\omega)\, e^{-i\omega t}\, ,
\label{corrfunc}
\end{eqnarray}
where 
\begin{equation}
\Im m \,G_{ij}^{\sigma}(\omega) = \frac{1}{2i}\, \left[G_{ij}^{\sigma}(\omega)-
G_{ji}^{\sigma *}(\omega)\right]\, .
\end{equation}

After Fourier transforming equation \ref{chiHF} and using equation 
\ref{corrfunc} one obtains

\begin{equation}
\chi_{ij}^{HF}(\omega) = -\frac{1}{\pi}\int d\omega'f(\omega')\, 
\left[\Im m\, G_{ji}^{\up}(\omega')\, G_{ij}^{\down}(\omega'+\omega)+
\Im m\, G_{ij}^{\down}(\omega')\, G_{ji}^{\up -}(\omega'-\omega)\right]\, ,
\label{chiHFfin}
\end{equation}
where $G_{ji}^{\sigma -}(\omega)$ is the advanced one-particle Green function.

Since we are interested in multilayer structures having translational symmetry
parallel to the layers, it is convenient to work with a mixed representation 
by choosing our basis as Bloch sums in a single atomic plane $\ell$ defined by 
\begin{equation}
\phi_\ell(\vec{q_\parallel})=\frac{1}{N_\parallel}\sum_{j\in\ell}
\varphi(\vec{R_j})\, e^{i\vec{q_\parallel}\cdot\vec{R_j}}\, .
\end{equation}
Here $\varphi(\vec{R_j})$ denotes an atomic orbital centered at site
$\vec{R_j}\in\ell$, $\vec{q_\parallel}$ is a wave-vector parallel to the 
layers, and $N_\parallel$ is the number of sites in this plane. Owing to the
fact that $\vec{q_\parallel}$ is a good quantum number, the 
Hartree-Fock susceptibility in such representation is given by
\begin{eqnarray}
\chi_{\ell \ell'}^{HF}(\vec{q}_{\parallel},\omega)=-\frac{1}{\pi}\int d\omega'
f(\omega') \frac{1}{N_{\parallel}}\sum_{\vec{k}_{\parallel}}\left[\Im m\, 
G_{\ell' \ell}^{\up}(\vec{k}_{\parallel},\omega')\, G_{\ell \ell'}^{\down}
(\vec{q}_{\parallel}+\vec{k}_{\parallel},\omega'+\omega) + \right.
\nonumber\\ 
\left. \Im m\, G_{\ell \ell'}^{\down}(\vec{k}_{\parallel},\omega')\, G_{\ell' 
\ell}^
{\up -}(\vec{k}_{\parallel}-\vec{q}_{\parallel},\omega'-\omega)\right]\, . 
\label{chiHFfin2}
\end{eqnarray} 
Similarly to equation \ref{fin}, the in-plane dynamic susceptibility 
satisfies the following equation:
\begin{equation}
\chi^{+-}_{\ell \ell}(\vec{q}_{\parallel},\omega) = 
\chi^{HF}_{\ell \ell}(\vec{q}_{\parallel},\omega) - \sum_m
\chi^{HF}_{\ell m}(\vec{q}_{\parallel},\omega)U_m\chi^{+-}_{m\ell}
(\vec{q}_{\parallel},\omega)\, .
\label{chiinplane}
\end{equation}

It is noteworthy that equation \ref{chiinplane} couples the susceptibility
matrix elements involving atomic planes with $U\neq 0$ only. Thus, for a
magnetic film of finite thickness on a non-interacting substrate, the set of
equations \ref{chiinplane} can be solved in matrix form as
\begin{equation}
\chi^{+-}(\vec{q}_{\parallel},\omega) = 
\left[I + \chi^{HF}(\vec{q}_{\parallel},\omega) U\right]^{-1}
\chi^{HF}(\vec{q}_{\parallel},\omega)\, ,
\label{chiinplane2}
\end{equation} 
where $\left[I + \chi^{HF} U\right]$ is a matrix in plane indices having 
finite dimension equal to the number of atomic planes of the magnetic film.

\section{Spin waves in some overlayers}

We consider a monolayer of a metallic ferromagnet on a surface of a
non-magnetic semi-infinite metallic substrate. We examine overlayers placed on
$(100)$ and $(110)$ surfaces of a simple cubic lattice. The electronic
structure of the system is described by the Hamiltonian given by equation
\ref{Hub-H}. We take into account hopping between nearest-neighbor sites only,
and assume that it is the same ($t_{ij}=t$) both in the substrate and in the
overlayer. We set the atomic energy levels $\epsilon_i$ and the effective
on-site Coulomb interactions $U_i$ both equal to zero in the substrate, and
fix the Fermi energy at $E_F=0.15$, which gives a bulk substrate occupancy of
$n=1.06$ electrons/atom. Here all energies are measured in units of the
nearest-neighbor hopping $t$. The HF ferromagnetic ground state of the system
is calculated self consistently. We choose $U_i=12$ in the surface layer, and
determine its atomic energy level $\epsilon_s$ so that the overlayer has
electronic occupancy of $n_s=1.68$ electrons/atom (appropriate for Co in this
single-band model). A relatively large value of $U$ in the overlayer was
chosen to guarantee a stable ferromagnetic HF ground state with a number of
holes in the majority-spin band of the overlayer much smaller than in the
minority one. The spin wave spectrum is obtained by calculating the surface
transverse spin susceptibility $\chi^{+-}(\vec{q}_{\parallel},\omega)$, using
equations \ref{chiinplane2} and \ref{chiHFfin2}.

First we study a ferromagnetic monolayer on a $(100)$ surface. Figure 1 shows
${\rm Im}\chi^{+-}(\vec{q}_{\parallel},\omega)$ calculated as a function of
energy $E=\hbar\omega$ for several values of $\vec{q}_{\parallel}$ along the
$[100]$ direction in the surface plane. The lifetime of a spin wave with wave
vector $\vec{q}_{\parallel}$ is inversely proportional to the width of the
peak of ${\rm Im}\chi^{+-}(\vec{q}_{\parallel},\omega)$. Such width may be
influenced by the small imaginary part $\eta$ usually added to the energy in
numerical calculations of $\chi^{+-}(\vec{q},\omega)$. Our calculations for
the $[100]$ direction were all made with the value of $\eta=1 \times 10^{-2}$,
and for the $[110]$ direction we have used $\eta=5 \times 10^{-3}$ for
numerical convergency reasons. It is clear from figure 1 that the spin waves
in the overlayer become strongly dumped for increasing values of
$q_{\parallel}$. The spin wave energies $E(\vec{q}_{\parallel})$ were obtained
from the position of the peak in ${\rm
  Im}\chi^{+-}(\vec{q}_{\parallel},\omega)$. The inset in figure 1 shows that
E varies quadratically with $q_\parallel$ over a wide range of values in the
first Brillouin zone.

The explanation for the spin wave dumping in the overlayer, given in
references [\onlinecite{phan}], is based on the tunneling of holes from the
non-magnetic substrate into the majority-spin band of the overlayer.  Without
a substrate, the free-standing ferromagnetic monolayer would have a
well-defined Stoner gap, and the spin waves infinitely long lifetimes. The
tunneling of holes from the substrate into the overlayer destroys this
well-defined Stoner gap allowing the spin waves to decay into electron-hole
pairs.  To prove that such explanation is correct we have gradually
disconnected the overlayer from the substrate by artificially reducing the
hopping $t_{\perp}=\alpha t$ between the surface layer and the substrate. We
consider several values of $0\leq\alpha\leq 1$, recalculating in each case the
HF ground state self consistently. The corresponding values of the surface
layer magnetic moments are listed in table 1. As expected, the ground state
magnetic moment slightly increases as $\alpha$ decreases. Figure 2 shows ${\rm
  Im}\chi^{-+}(E,q_{\parallel})$, calculated as a function of $E$ for
different values of $\alpha$, and $q_{\parallel}= 0.083\times 2\pi/a$, where
$a$ is the lattice constant.  When $\alpha$ decreases so does the tunneling of
holes from the substrate into the overlayer.  Thus, the decaying probability
of the spin waves reduces and their lifetimes increase. Consequently, the
widths of the spin wave peaks become narrower, as evidenced in figure 2. It is
also noticeable from figure 2 that the spin wave energies become smaller as
$\alpha$ decreases. This is partially due to the reduction of the exchange
coupling between the local moments mediated by the substrate when $\alpha$
decreases. This shows that the presence of a non-magnetic substrate may
substantially affect the spin wave spectrum of the monolayer.

\begin{table}
\begin{center}
\begin{tabular}{|c||c|c|c|c|c|c|}
\hline
$\alpha$ & 1.0 & 0.8 & 0.6 & 0.3 & 0.1 & 0.0\\ \hline
$m_s$ & 0.275 & 0.292 & 0.304 &  0.316 &  0.320 &  0.320 \\ \hline
\end{tabular}
\caption{Magnetic moments of the surface layer (in units of the Bohr magneton)
  calculated for different values of the hopping $t_{\perp}=\alpha t$ between
  the substrate and the overlayer. Except for $\alpha=0$ all values of $m_s$
  have been determined self-consistently.}
\end{center}
\end{table}

We now examine a ferromagnetic monolayer on a $(110)$ surface. Figures 3 and 4
show results of ${\rm Im}\chi^{+-}(\vec{q}_ {\parallel},E)$, calculated as a
function of energy for several values of $\vec{q}_{\parallel}$, along the
$[001]$ $(\hat{z})$ and $[1\bar{1}0]$ $(\hat{\xi})$ directions in the surface
plane, respectively. The spin wave energy for the $(110)$ surface is not
isotropic in $\vec{q_{\parallel}}$-space. The origin for such anisotropy lays
on the crystalline structure of the $(110)$ overlayer.  For a simple cubic
lattice it is formed by chains of nearest-neighbor sites along the $\hat{z}$
direction that in the absence of second-neighbor hopping are linked to each
other via the substrate only.  Thus, without a substrate (i.e., for a
free-standing $(110)$ monolayer), those chains would be uncoupled, and no
energy would be required to excite long wavelength spin waves propagating
perpendicularly to the chains. \cite{Tovar} The inset in figure 3 shows that
the dispersion relation for spin waves propagating along $\hat{z}$ varies
quadratically with $q_z$ over a wide range of values in the first Brillouin
zone. In contrast, the inset of figure 4 shows that for $\vec{q_\parallel}$
perpendicular to the chains, $E(q_{\xi})$ deviates from the quadratic behavior
for relatively low values of $q_{\xi}$.

By comparing the insets of figures 3 and 4, we note that the energy of a spin
wave propagating along $[1\bar{1}0]$ in the $(110)$ overlayer is smaller than
when it propagates along $\hat{z}$-direction with the same
$\left|\vec{q_\parallel}\right|$. The difference in energies, however, is not
as large as previously found for unsupported monolayers.\cite{Tovar} The
reason is the exchange interaction between the chains, that is mediated the
substrate. Although smaller, it is comparable to the direct exchange
interaction between nearest-neighbor sites along the chains, leading to a
stiffness along $[1\bar{1}0]$ which is of the same order of magnitude of that
along $\hat{z}$.

By reducing the hopping from the overlayer to the substrate, the inter-chain
coupling decreases and the energy to excite a spin wave propagating along
$[1\bar{1}0]$ becomes substantially smaller. This is illustrated in figure 5,
and it is qualitatively in accordance with what has been found in reference
\onlinecite{Tovar}.

\section{Conclusions}

We have investigated the influence of a non-magnetic metallic substrate on the
spin excitations in ultrathin ferromagnetic overlayers. Both the spin wave
energies and lifetimes have been determined by calculating the transverse
dynamic spin susceptibility $\chi^{+-}(\vec{q_{\parallel}},E)$ for different
surface crystalline orientations. We have found that the spin waves in the
overlayer is strongly dumped due to the presence of the substrate. By
gradually reducing the hopping between the substrate and the overlayer we
verify that such dumping is caused by the tunneling of holes from the
substrate into the overlayer, as previously pointed out by Mathon et
al.\cite{phan}.  For the (110) overlayer we have found that the spin wave
energy depend upon the direction of the wavevector $\vec{q_\parallel}$ with
which it is excited.  Such dependence, however, is much less pronounced than
previously obtained for unsuported monolayers. We argue that this is due to
the enhancement of the exchange coupling between the local moments in the
overlayer mediated by the substrate that strongly affects the spin wave
energies.

\section{acknowledgments}

We wish to thank J. d'Albuquerque e Castro for very helpful discussions.
Financial support from CNPq, FAPEMIG and FAPERJ (Brazil) is gratefully 
acknowledged.

\newpage

\begin{figure}
\caption{ Spin wave spectrum for the $(100)$ overlayer. The figure shows 
  ${\rm Im}\chi^{+-}(\vec{q}_\parallel,\omega)$ calculated as a function of
  energy $E=\hbar\omega$ for several values of $\vec{q}_\parallel = q_x
  \hat{x}$ along the $[100]$-direction. Solid line is for $q_x=0.042$, dashed
  line for $q_x=0.063$, dot-dashed line for $q_x=0.083$, and long-dashed line
  for $q_x=0.166$. All values of $q_x$ are in units of $2\pi/a$ where $a$ is
  the lattice constant. The inset show the corresponding spin wave energies
  obtained from the positions of the peaks of ${\rm Im}\chi^{+-}$.}
\label{fig1}
\end{figure}

\begin{figure}
\caption{ ${\rm Im}\chi^{-+}(q_{\parallel},E)$, calculated as a function of 
  $E=\hbar\omega$, for a fixed $q_{\parallel}= 0.083$ along the $[100]$
  direction, and different values of the hopping $t_{\perp}=\alpha t$ between
  the $(100)$ surface layer and the substrate. Thin-solid line is for
  $\alpha=1.0$, dotted line for $\alpha=0.8$, dashed line for $\alpha=0.6$,
  long-dashed line for $\alpha=0.3$, dot-dashed line for $\alpha=0.1$, and
  thick-solid line for $\alpha=0$}
\label{fig2} 
\end{figure}

\begin{figure}
\caption{Spin wave spectrum for the $(110)$ overlayer. The figure shows 
  ${\rm Im}\chi^{+-}(\vec{q}_\parallel,\omega)$ calculated as a function of
  energy $E=\hbar\omega$ for several values of $\vec{q}_\parallel = q_z
  \hat{z}$ along the $[001]$-direction. Solid line is for $q_z=0.1$, doted
  line for $q_z=0.2$, dashed line for $q_z=0.3$, and long dashed line
  for $q_z=0.4$. All values of $q_z$ are in units of $2\pi/a$ where $a$ is
  the lattice constant. The inset show the corresponding spin wave energies
  obtained from the positions of the peaks of ${\rm Im}\chi^{+-}$}
\label{fig3} 
\end{figure} 

\begin{figure} 
\caption{Spin wave spectrum for the $(110)$ overlayer. The figure shows 
    ${\rm Im}\chi^{+-}(\vec{q}_\parallel,\omega)$ calculated as a function of
    energy $E=\hbar\omega$ for several values of $\vec{q}_\parallel = q_\xi
    \hat{\xi}$ along the $[1\bar{1}0]$ direction. Solid line is for
    $q_\xi=0.1$, dashed line for $q_\xi=0.2$, and long dashed line for
    $q_\xi=0.3$. All values of $q_\xi$ are in units of $2\pi/a$ where $a$
    is the lattice constant. The inset show the corresponding spin wave
    energies obtained from the positions of the peaks of ${\rm Im}\chi^{+-}$.}
\label{fig4} 
\end{figure} 

\begin{figure} 
\caption{${\rm Im}\chi^{-+}(q_{\parallel},E)$, calculated as a function of 
  $E=\hbar\omega$, for a fixed value of $\vec{q}_\parallel = 0.2\hat{\xi}$,
  where $\hat{\xi}$ is a unit vector along the $[1\bar{1}0]$-direction in the
  $(110)$ surface plane.  The solid and dashed lines correspond to different
  values of the hopping $t_{\perp}=\alpha t$ between the $(110)$ surface layer
  and the substrate. Solid line is for $\alpha=1.0$, and dashed line for
  $\alpha=0.1$.}
\label{fig5} 
\end{figure}

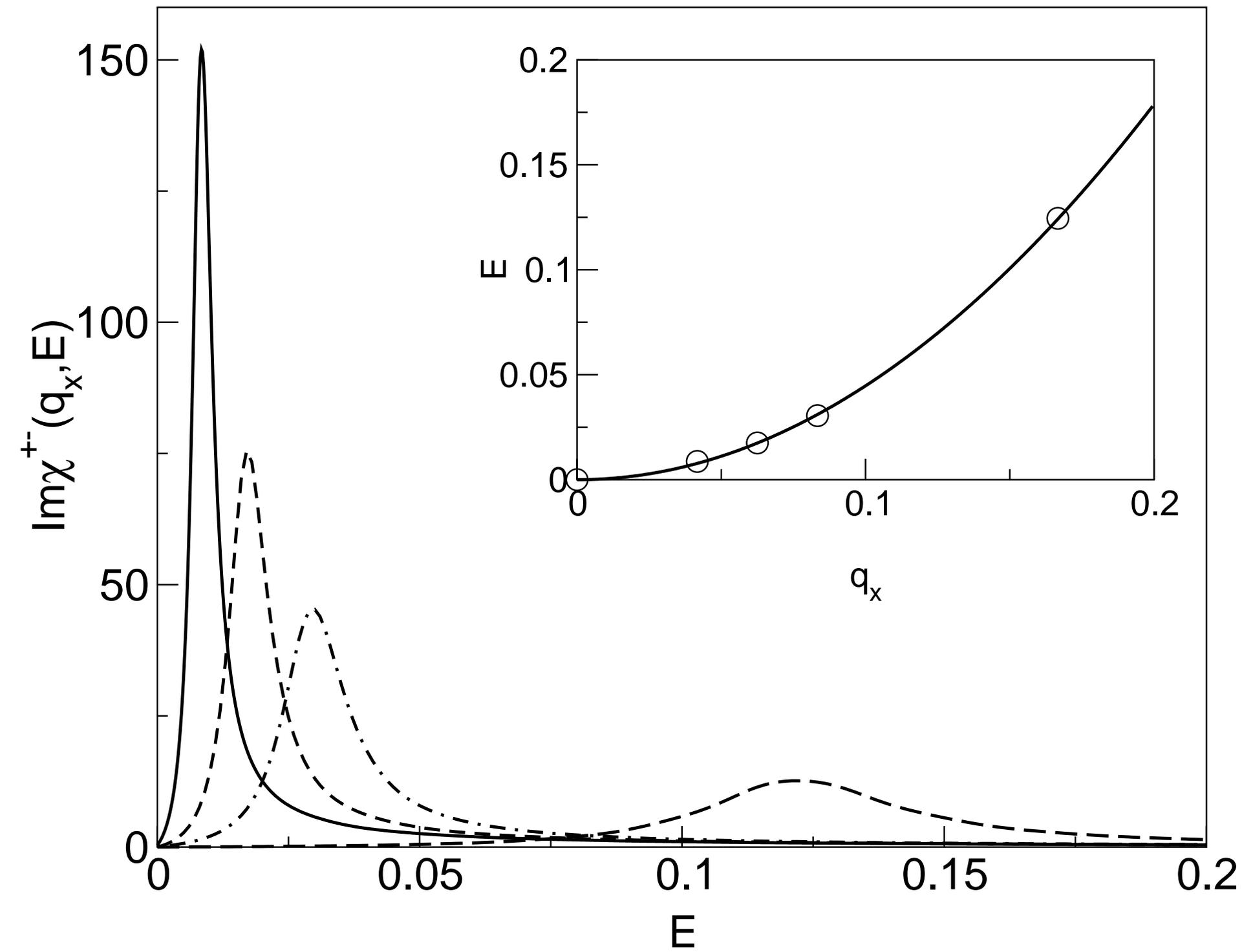
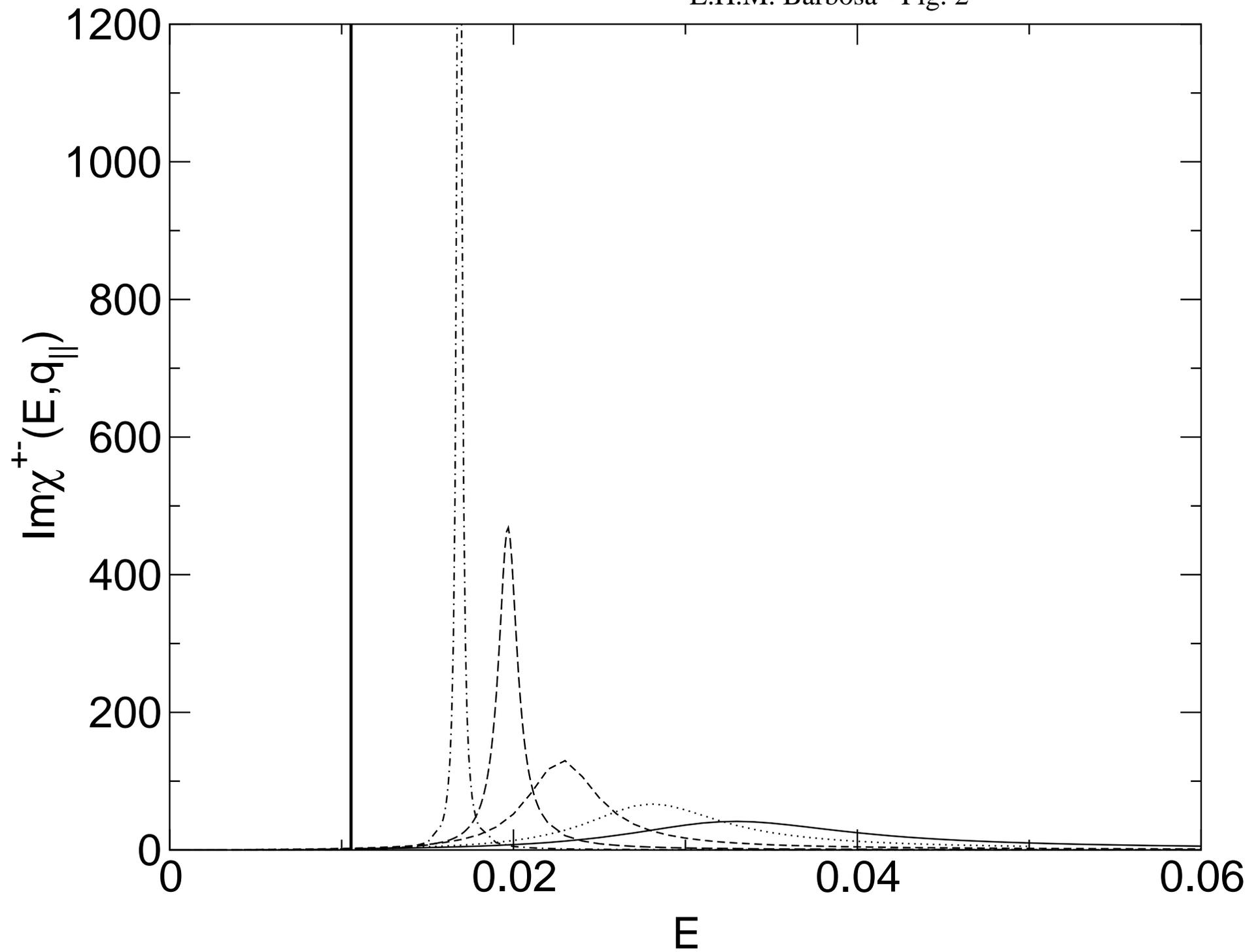
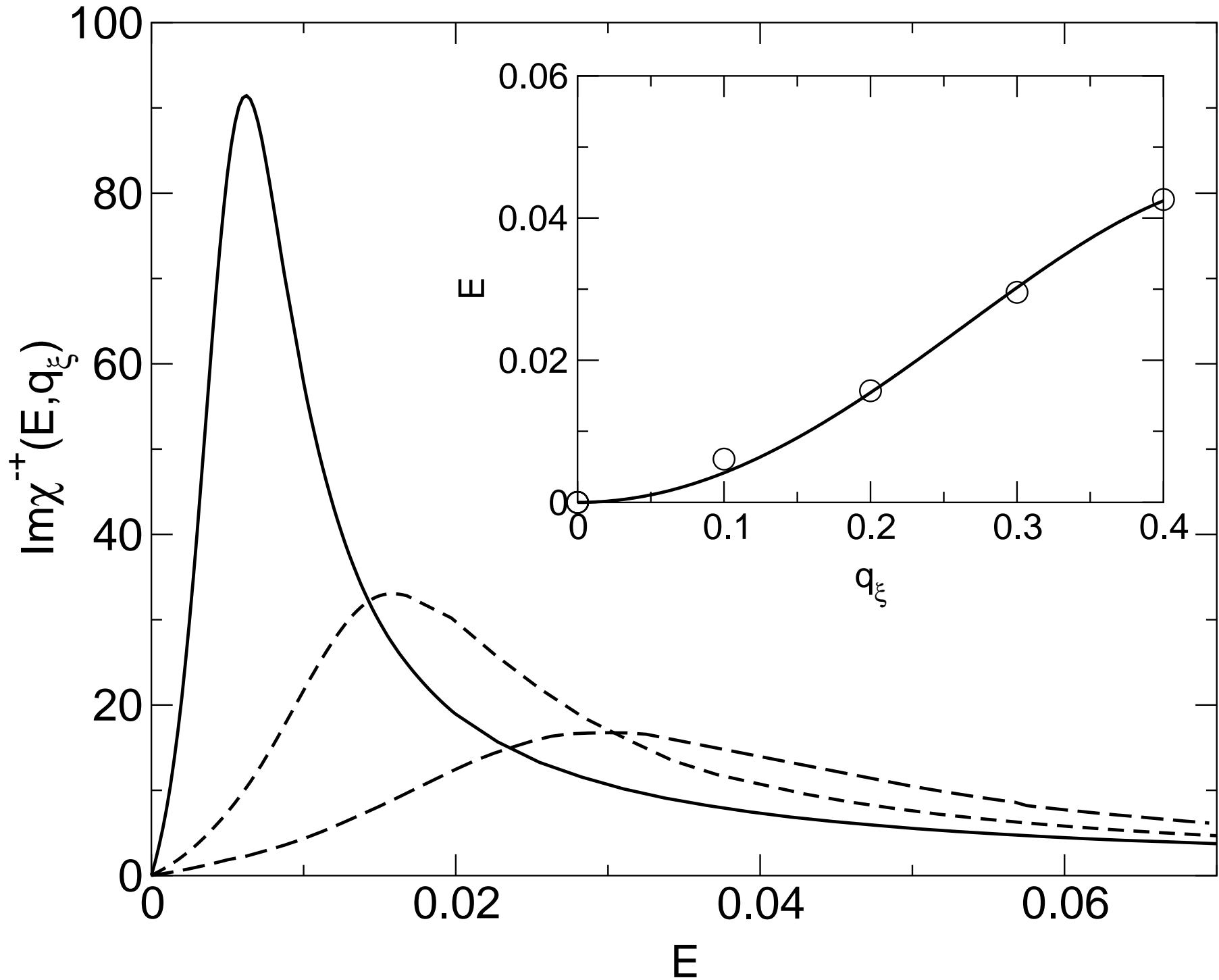
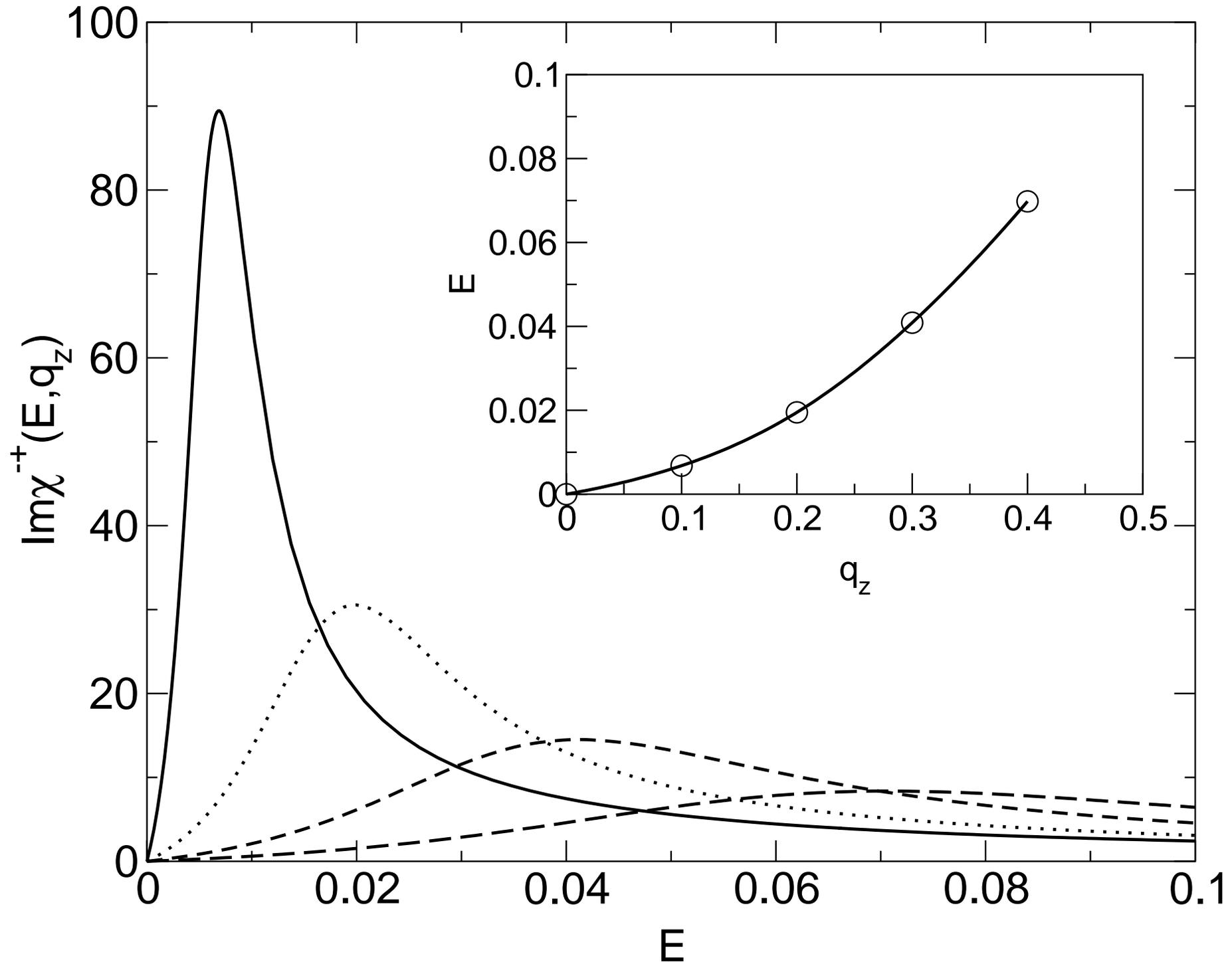
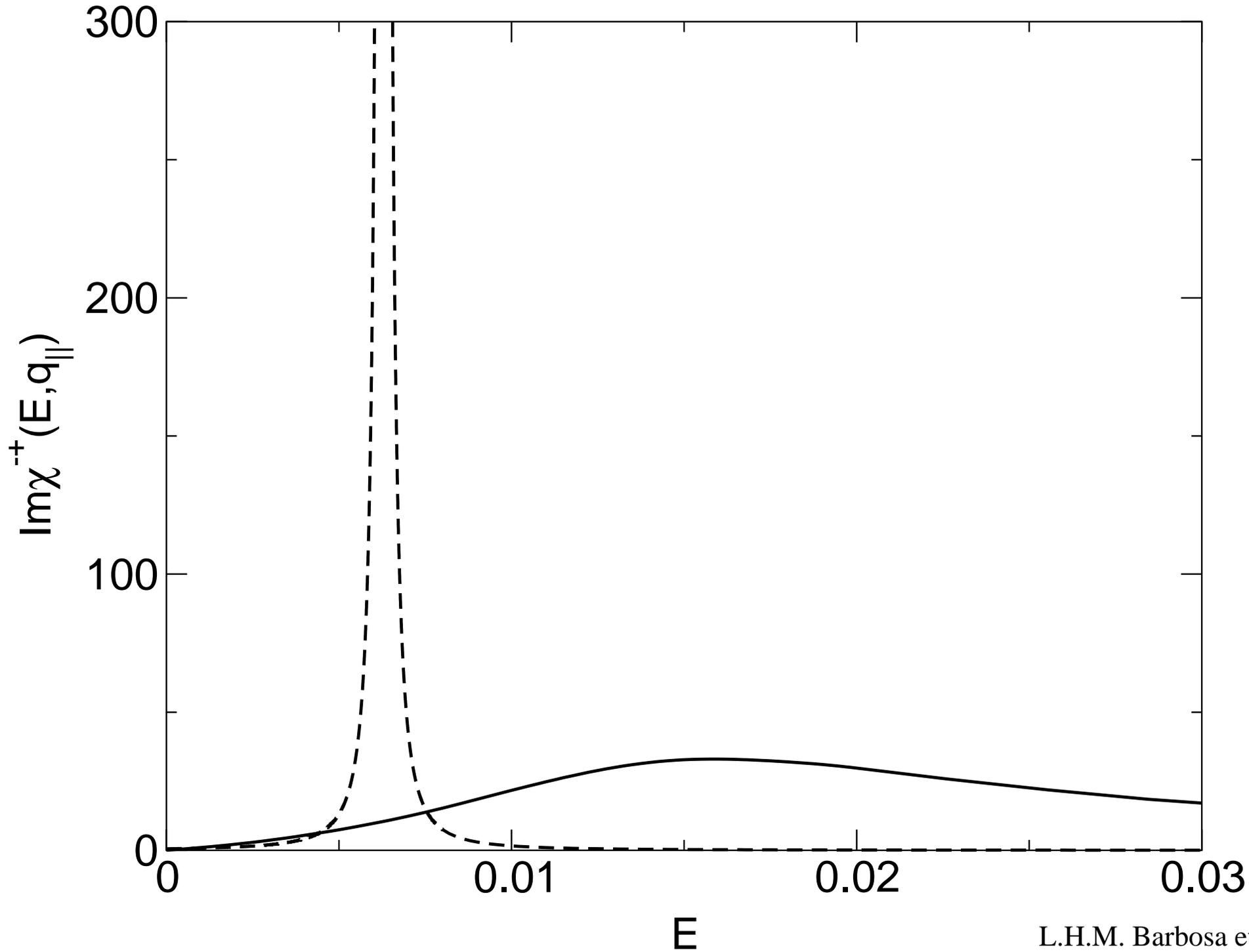
\end{document}